\documentclass[aps]{revtex4} 
\usepackage{times}
\usepackage{graphicx}
\begin{document}

\title{Accurate and efficient description of protein vibrational
dynamics: comparing molecular dynamics and Gaussian models}

\date{\today}
 
\author{Cristian Micheletti, Paolo Carloni and Amos Maritan\\
\small International School for Advanced Studies (S.I.S.S.A.) and INFM, Via Beirut 2-4, 34014 Trieste, Italy
}
\date{\today}

\begin{abstract} 
Current all-atom potential based molecular dynamics (MD) allow the
identification of a protein's functional motions on a wide-range of
time-scales, up to few tens of ns. However, functional large scale motions
of proteins may occur on a time-scale currently not accessible by
all-atom potential based molecular dynamics. To avoid the massive
computational effort required by this approach several simplified
schemes have been introduced. One of the most satisfactory is the
Gaussian Network approach based on the energy expansion in terms of
the deviation of the protein backbone from its native
configuration. Here we consider an extension of this model which
captures in a more realistic way the distribution of native
interactions due to the introduction of effective sidechain
centroids. Since their location is entirely determined by the protein
backbone, the model is amenable to the same exact and computationally
efficient treatment as previous simpler models. The
ability of the model to describe the correlated motion of protein
residues in thermodynamic equilibrium is established through a series
of successful comparisons with an extensive (14 ns) MD simulation
based on the AMBER potential of HIV-1 protease in complex with a
peptide substrate. Thus, the model presented here emerges as a
powerful tool to provide preliminary, fast yet accurate
characterizations of proteins near-native motion.
\end{abstract}

\maketitle

\section{Introduction}

Considerable insight into the biological activity of a protein can be
gained by identifying its large-scale functional movements. Ideal
tools for a detailed characterization of such dynamical properties are
constituted by computational techniques such as molecular dynamics
(MD) simulations based on effective all-atom potentials
\cite{Karplus_ACR}. By these means it is possible, at present, to
follow numerically the dynamical evolution of a protein of a few hundred
residues in its surrounding solvent over time intervals of tens of
nanoseconds.

Such time-scales allow to gain considerable insight into important
aspects of protein dynamics and to make quantitative connections with
experimental quantities such as NMR order parameters \cite{nmr1,nmr2}
and Trp fluorescence spectra \cite{karplus_book}. Other complex
conformational changes are however difficult or impossible to be
observed. Examples include protein-protein molecular recognition,
rearrangements occuring upon ligand binding etc. which all involve
time-scales of the order of 1 $\mu$s or longer\cite{go82}.  In
addition, the simulated trajectory might not be sufficiently long that
thermodynamic averages can be legitimately replaced with dynamical
ones\cite{Hess2002}.

Several studies have attempted to bridge the gap between the time
scales of feasible MD simulations and the ones of
biologically-relevant protein motion by recoursing to a mesoscopic
rather than a microscopic approach\cite{tir96}. In fact, the
large-scale dynamical features encountered in MD trajectories can be
conveniently interpreted, at a first approximation, as a superposition
of independent harmonic modes \cite{go82}. This
observation was complemented by Tirion who pointed out that, in a
normal mode analysis of protein vibrations, the detailed classical
force-field could be replaced by suitable harmonic couplings with the
same spring constants \cite{tir96}. These results stimulated a variety
of studies where the elastic properties of proteins were described
through coarse-grained models where amino acids are replaced by
effective centroids corresponding to the $C_\alpha$ atoms and the
energy function is reduced to harmonic couplings between pairs of
spatially close centroids. These approaches, in particular the
Gaussian and anisotropic network models (GNM and ANM), have been found
to be in accord with both experimental and MD results
\cite{bah97,doruker2000,ani01}. 

Here we introduce an extended network gaussian model which, at
variance with previous approaches, incorporates effective $C_\beta$
centroids ``tethered'' to the $C_\alpha$ atoms. The presence of the
effective $C_\beta$'s allows a good control of the directionality of
pairwise interactions in the protein and thus leads to an improved
vibrational description. Furthermore, the fact that the sidechains
degrees of freedom are entirely controlled by the $C_\alpha$'s has the
crucial implication that the computational effort required to
characterize the system is exactly the same as for models based only
on $C_\alpha$'s.

The model equilibrium dynamics is compared against MD simulations
based on all-atom effective potentials. We provide several
quantitative estimates for how and in what sense, the simplified
quadratic approaches can provide a complement of the more accurate but
also much more computationally-demanding all-atom potential based MD
calculations.   From a general point of view, the
ideal term of reference would be constituted by a direct experimental
determination of the quantities of interest here, such as the
correlation of residues' motion. However, since this is not presently
feasible, such detailed information can only be obtained from MD
simulations. Although such simulations do not provide an absolute term
of comparison, it is certainly an adequate reference for the much
simplified protein models and energy functionals considered here.

\begin{figure}
\includegraphics[width=3.0in]{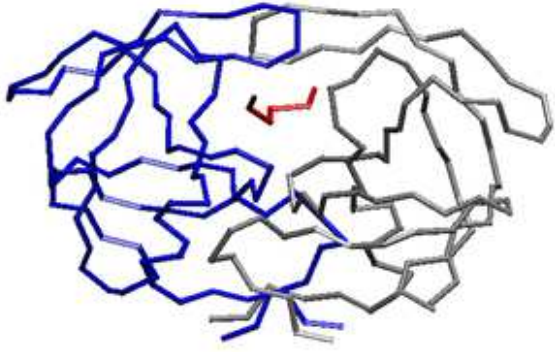}
\caption{Backbone trace of the complex formed by the HIV-1 protease
homodimer and the bound substrate. Each monomer is composed of 99
residues; The model substrate consists of six amino acids
\protect{\cite{piana2002,piana2002b}}.}
\label{fig:hiv}
\end{figure}

The reference system considered here is the complex formed by the
HIV-1 protease dimer with a bound model substrate
(Fig. \ref{fig:hiv}).  This protein appears to be suitable to serve as
reference for our model calculations in many respects. First, its
dynamics in aqueous solution has been extensively investigated over
more than 10 ns by all-atom effective potentials MD simulation
\cite{piana2002,piana2002b}. Second, a large-scale motion analysis
based on this MD simulation has been already performed. Third, this
protein/substrate complex does not contain metal ions, prostetic
groups, cofactors or non standard amino acids, and therefore it is
amenable as a first test system for our model calculations, which
consider only standard amino acids. Finally, this protein is of
outstanding pharmaceutical relevance (it is one of the two targets
currently used in anti-AIDS therapy) and its dynamics has been
revealed as a key ingredient for the enzymatic function and for
rationalizing resistance data \cite{condra,patick,piana2002,pcr}.

The data obtained from the MD simulation of
refs. \cite{piana2002,piana2002b} have been used here to evaluate all
the MD-related quantities used for comparison against the results for
quadratic models. In support of the general applicability of Gaussian
approaches to capture the details of proteins essential motion we also
report a comparison of the model predictions against data from a
recent MD study of the NGF-trkA complex formed by the nerve growth
factor and the tyrosin Kinase A receptor\cite{settanni_ngf}.

\section{Theory}

\subsection{The model}

The starting point of the present analysis is the expansion of the
Hamiltonian in terms of the deviations of the amino acids from
their reference native positions. The underlying
assumption is that a protein immersed in aqueous solution vibrates
around its native state with amplitudes so small to justify the
quadratic expansion around the minimum of the potential energy
function. {\em A posteriori} this does not seem to be a drastic
restriction \cite{go82,go_gauss91,doruker2000} despite both the
large-amplitude of motion observed in dynamical trajectory and the
existence of several native sub-states, instead of a unique energy
minimum \cite{substates}.

To reduce the spatial degrees of freedom of a protein we adopt a
coarse-grained model where a two-particle representation is used for
each amino acid: besides the $C_\alpha$ atom, an effective $C_\beta$
centroid is employed to capture, in the simplest possible way, the
sidechain orientation in a given amino acid (except for GLY for which
only the $C_\alpha$ atom is retained). To distinguish the proposed
model from those based on the $C_\alpha$ only representation will
shall refer to it as the $\beta$ Gaussian model ($\beta$GM for
brevity).  Since our focus is to study the concerted vibration of
various amino acids around the native state of the system, the
Hamiltonian that is adopted incorporates, accordingly, pairwise
interactions between all pairs of particles that are sufficiently
spatially close in the native state.  Formally, the system energy
function evaluated on a trial structure, $\Gamma$, takes on the form

\begin{equation}
{\cal H} (\Gamma)  = {\cal H}_{BB} (\Gamma) + {\cal H}_{\alpha\alpha} (\Gamma) 
+ {\cal H}_{\alpha\beta} (\Gamma) + {\cal H}_{\beta\beta} (\Gamma) 
\label{eqn:ham}
\end{equation}

\noindent where 
\begin{eqnarray}
{\cal H}_{BB} (\Gamma) &=&  k \sum_i V^{CA-CA} (d_{i,i+1}^{CA-CA})\nonumber \\
{\cal H}_{\alpha\alpha} (\Gamma) &=& \sum_{i<j} \Delta^{CA-CA}_{ij} V^{CA-CA} (d_{i,j}^{CA-CA}) \nonumber\\
{\cal H}_{\alpha\beta} (\Gamma) &=& \sum_{i,j} \Delta^{CA-CB}_{ij} V^{CA-CB} (d_{i,j}^{CA-CB}) \nonumber\\
{\cal H}_{\beta\beta} (\Gamma) &=& \sum_{i<j} \Delta^{CB-CB}_{ij} V^{CB-CB} (d_{i,j}^{CB-CB}), 
\label{eqn:hamb}
\end{eqnarray}

\noindent In expressions (\ref{eqn:hamb}) $\Delta^{XY}_{ij}$ is the
native contact matrix that takes on the values of 1 [0] if the native
separation of the effective particles of type $X$ and $Y$, belonging
respectively to residues $i$ and $j$, is below [above] a certain
cutoff value, $R$. With $d^{XY}_{ij}$, on the other hand, we denote
the actual separation of the particles in the trial structure,
$\Gamma$. The indices $i$ and $j$ run over all integer values ranging
from 1 up to the protein length, $N$. In particular, the interaction
between particles in consecutive amino acids, $| i-j| =1$, leads to a
simple treatment of the protein chain connectivity. However, to
account for the much higher strength of the peptide bond with respect
to non-covalent contact interactions between amino acids, we have
added in (\ref{eqn:ham}) an explicit chain term, ${\cal H}_{BB}$, where
the interaction of consecutive $C_\alpha$'s is controlled by $k >0$.

By construction, the minimum of the various interaction terms is
attained for the native separation of each pair of particles. This
ensures that the native state is at the global energy minimum.  For
small fluctuations around the native structure, the potential
interaction energy of two particles, $i$ and $j$, can be expanded in
terms of the deviations from the native distance-vector,
$\vec{r}_{ij}$. If we indicate the deviation vector as $\vec{x}_{ij}$,
so that the total distance vector is $\vec{d}_{ij} = \vec{r}_{ij} +
\vec{x}_{ij}$, we can approximate the pairwise interaction as

\begin{equation}
V(d_{ij}) \approx 
V(r_{ij}) + { V^{\prime \prime}(r_{ij}) \over 2} \sum_{\mu,\nu}
{r_{ij}^\mu\, r_{ij}^\nu \over r^2_{ij}} x_{ij}^\mu\,
x_{ij}^\nu 
\label{eqn:vexp}
\end{equation}

\noindent where $\mu$ and $\nu$ denote the cartesian components, $x$,
$y$ and $z$, and $V^{\prime \prime}$ is the second derivative of $V$.
Several models have been introduced previously, where the quadratic
expansion (\ref{eqn:vexp}) was used in a context where only
interactions among $C_\alpha$ centroids where considered, such as in
the anisotropic gaussian model (ANM), recently introduced to study the
vibrational spectrum of proteins \cite{ani01}.

Based on this quadratic expansion the Hamiltonian of eq. (\ref{eqn:ham})
can be approximated as,

\begin{equation}
\tilde {\cal H} = { 1 \over 2}\, \sum_{ij,\mu\nu} x^{CA}_{i,\mu}
M^{CA-CA}_{ij,\mu \nu} x^{CA}_{j,\nu}\ ,
\label{eqn:ham2}
\end{equation}

\noindent where ${M}$ is a $3N$x$3N$ symmetric matrix. The elastic
response of the system is uniquely dictated by the eigenvalues and
eigenvectors of ${M}$.

What differentiates the $\beta$GM from several previous studies is the
presence of the interactions between $C_\alpha$ and $C_\beta$ and
$C_\beta$-$C_\beta$ (besides the extra strength of the chain term).
The introduction of the $C_\beta$ centroids in the protein description
leads, in principle, to a more complicated Hamiltonian, with the
additional $C_\beta$'s degrees of freedom:

\begin{eqnarray}
{\cal H} = &&{1 \over 2} \sum_{ij,\mu\nu}  x^{CA}_{i,\mu} M^{CA-CA}_{ij,\mu\nu}
x^{CA}_{j,\nu} \nonumber \\
&&+ \sum_{ij,\mu\nu}  x^{CA}_{i,\mu} M^{CA-CB}_{ij,\mu \nu}
x^{CB}_{j,\nu} \nonumber \\
&&+ {1 \over 2} \sum_{ij,\mu\nu}  x^{CB}_{i,\mu} M^{CB-CB}_{ij,\mu \nu}
x^{CB}_{j,\nu} \ .
\label{eqn:ham3}
\end{eqnarray}

\noindent However, the location of the $C_\beta$ atoms in a protein
structure is almost uniquely specified by the geometry of the peptide
chain. An accurate method that predicts the location of the $C_\beta$
atoms from the CA trace of a protein is the geometric construction of
Park and Levitt\cite{cb_construct}, which assigns the $i$th $C_\beta$
location given the positions of the $C_\alpha$'s of residues $i-1$,
$i$ and $i+1$, allows to place the fictitious $C_\beta$ at a distance
of 0.3 \AA\ from the crystallographic location. Such excellent
agreement clarifies that the degrees of freedom of the $C_\beta$
centroids should not be considered independent from the $C_\alpha$
ones. On the contrary, the $C_\beta$'s can be viewed as rigidly
``tethered'' to the $C_\alpha$ and hence the fluctuations of the
former are dictated by those of the latter.

Although in principle one could use the original rule of Park and
Levitt\cite{cb_construct}, we have adopted a simpler construction
scheme which places the $C_\beta$ exactly in the plane specified by
the local $C_\alpha$ trace. This simplifies the construction of the
$M$ matrices which remain ``diagonal'' in the cartesian components
(e.g. the $x$ component of the reconstructed $C_\beta$ depends only on
the $x$ components of the neighbouring $C_\alpha$'s. More precisely,
the location of the $i$th $C_\beta$ is given by

\begin{equation}
\vec{r}_{CB}(i) = \vec{r}_{CA}(i) + l {2 \, \vec{r}_{CA} (i) -
\vec{r}_{CA} (i+1) -\vec{r}_{CA} (i-1) \over | 2 \, \vec{r}_{CA} (i) -
\vec{r}_{CA} (i+1) -\vec{r}_{CA} (i-1)|}
\end{equation}

\noindent For reasons of self-consistency of the model, this
construction rule is used to determine the contact matrices involving
the effective $C_\beta$ centroids that are used in place of those
depending on the crystallographic $C_\beta$ locations in eqn
(\ref{eqn:hamb}). To leading order in the deviations of the
$C_\alpha$ atoms, the deviations of $r_{CB}(i)$ thus becomes:

\begin{equation}
\vec{x}_{CB}(i) \approx l { 2 \, \vec{x}_{CA} (i) - \vec{x}_{CA} (i+1)
-\vec{x}_{CA} (i-1)  \over | 2 \, \vec{r}_{CA} (i) - \vec{r}_{CA} (i+1)
-\vec{r}_{CA} (i-1)|} \ .
\label{eqn:cbfluc}
\end{equation}

\noindent where $l = 3$ \AA. By using this rule, one parametrizes, in
terms of the $C_\alpha$ positions the effective $C_\beta$ location of
all residues except for GLY and for the initial and final residues
which lack one of the flanking $C_\alpha$'s. When the resulting
expressions (\ref{eqn:cbfluc}) are substituted in equation
(\ref{eqn:ham3}) one obtains an effective quadratic Hamiltonian which,
as in equation (\ref{eqn:ham2}) involves only the $C_\alpha$
deviations but coupled through a new effective matrix, $\tilde{M}$
which is distinguished from previous matrices by a tilde
superscript. The book-keeping operations necessary to calculate the
elements of such matrix are conveniently implemented with the aid of a
computer. Thus, the computational cost and difficulty to characterize
the elastic response of the protein is reduced to exactly the same as
models with $C_\alpha$ atoms only.  In spite of the same computational
cost, the $\beta$GM appears to have several advantages in terms of the
ability to capture the low-frequency motion and other vibrational
properties of proteins, as will be seen below.

\subsection{Equilibrium properties}

The derivation of the vibrational properties of a protein from the
quadratic expansion of the Hamiltonian can be done, broadly speaking,
in two different ways: the normal mode analysis and the Langevin
analysis. What differentiates the two approaches is the view of the
role of the solvent on the system dynamics. If one assumes that the
motion of the protein is not significantly damped by the interaction
with the solvent, then the normal modes picture can be applied to
study the system dynamics around the native state by solving the
Newton's dynamical equations
\cite{levitt85,Karplus85,Case94,Hinsen98,go_gauss91,tirion93,hiv99,hal99}:

\begin{equation}
m_i\, \ddot{x}_{i,\mu}  = \sum_{j,\nu} \, \tilde{M}_{ij,\mu\nu}
x^{CA}_{j,\nu}
\label{eqn:m}
\end{equation}

\noindent The eigenfrequecies and eigenvectors are hence obtained by
diagonalizing a matrix derived from $\tilde{M}$ by an appropriate
mass-weighting \cite{Goldstein}.

Although the normal-mode analysis allows a straightforward dynamical
characterization it is of dubious applicability since protein motion
in a solvent does not resemble a superposition of pure harmonic
oscillations. In fact, several theoretical, experimental and
computational studies, have shown that the dynamics of a protein is
severely overdamped by the interaction with the solvent
\cite{Karplus76,Karplus82,Karplus85,Hinsen98}. The description of the
motion in terms of overdamped dynamics appears to be particularly
valid for the protein's low-frequency vibrations, that are the most
interesting ones due to their expected role in proteins functional
activities \cite{hal99}. This observation leads to the alternative
view of a heavily damped dynamics \cite{Howard}.

In this case, for small deviations from the reference positions, the
dynamics of the amino acids can be written as:

\begin{equation}
\dot{x}_{i,\mu} (t) = - \sum_{j,\nu} \tilde{M}_{ij,\mu\nu} \, x_{j,\nu}(t)
+ \eta_{i,\mu}(t)
\end{equation}

\noindent where the time unit has been implicitly chosen so that the
viscosity coefficients (assumed to be equal for all particles) are
set equal to 1 and the stochastic noise terms satisfy
\cite{chandrasekhar}:

\begin{eqnarray}
\langle \eta_{i,\mu} \rangle &=& 0\\
\langle  \eta_{i,\mu} \eta_{j,\nu} \rangle&=& \delta_{i,j}\, 
\delta_{\mu,\nu} 2 \kappa_B\, T \ .
\end{eqnarray}

\noindent These two conditions ensure, in the long run, the onset of
canonical thermal equilibrium, so that the equilibrium probability of
a given configuration, $\{x\}$, for the particles in the system is
controlled by the Boltzmann factor:

\begin{equation}
e^{-\beta {\cal H}(\{x\}) } = e^{ -{\beta}\, \sum_{ij,\mu\nu} x_{i,\mu} \tilde{M}_{ij,\mu \nu}
x_{j,\nu}}\ .
\label{eqn:boltz}
\end{equation}

In this case, no periodic motion of the system can exist in the
absence of external periodic excitations, since any structural
deformation will be dissipated by a damped dynamics. The
standard theory of stochastic processes \cite{chandrasekhar,levitt85}
shows that the eigenvalues of $\tilde{M}$ are inversely proportional
to the system relaxation times, and the corresponding eigenvectors
indicate the actual shape of the associated distortion of the system.

\subsection{Covariance matrices and temperature-factors}

Besides identifying the elementary modes of excitation of a protein,
it is important to calculate suitable thermodynamic quantities that
characterize the protein dynamics once thermal equilibrium with the
solvent has established. The main observable that can be calculated
within the gaussian model is the degree of correlation of the
displacement from the equilibrium (native position) of pairs of
$C_\alpha$'s. The thermodynamic average of the correlated
displacements, are easily obtained from the inversion of the ${M}$
matrix. In fact, after setting $1/\beta = K_B T =1 $, one has

\begin{equation}
\langle x_{i,\mu}\, x_{j,\nu} \rangle = {\tilde
M}^{-1}_{ij,\mu\nu}
\label{eqn:fullcij}
\end{equation}

\noindent where the brackets denote usual canonical thermodynamic
averages with the weight of eqn. (\ref{eqn:boltz}). The inverse
matrix, ${\tilde M}^{-1}_{ij,\mu\nu}$, is often referred to as
the covariance matrix. Since it provides directional details about the
correlated motion of pairs of residues we shall term it {\em full}
covariance matrix to distinguish it from the {\em reduced} one
discussed below which incorporates only a measure of the degree of
correlation (but no directional information).

The eigenvectors of the full covariance matrix represent the
three-dimensional independent modes of structural distortion for the
reference protein. The modes associated to the largest eigenvalues of
${\tilde M}^{-1}$ are the slowest to decay in a dissipative dynamics
and, hence, make the largest contribution to the mean-square
displacement of a given residues. The latter quantity is
straightforwardly calculated from eqn. (\ref{eqn:fullcij}):

\begin{equation}
\langle |\vec{x}_i |^2 \rangle = \sum_{\mu} {\tilde
M}^{-1}_{ii,\mu\mu}
\label{eqn:bfact}
\end{equation}

\noindent and can be directly connected to the temperature-factors
(also called B-factors) measurements reported in X-ray or
high-resolution NMR structural determinations \cite{normod}.

\noindent It is worth remarking that the full covariance matrix
provides information about the system elasticity not only in
conditions of isolation but also when an external force, $\vec{f}_i$,
is applied to a given residue, $i$. In fact, within the Gaussian
approximation, the average displacement of the $j$th amino acid from
its reference position due to the application of $\vec{f}_i$ is given
by

\begin{equation}
\langle x_{j,\nu} \rangle \propto \sum_{\mu} {\tilde
M}^{-1}_{ji,\nu\mu}\ f_{i,\mu}\ \ .
\end{equation}

\noindent As anticipated above, an important role in the analysis of
molecular dynamical trajectories is also played by the reduced
covariance matrix whose elements, $C_{ij}$, are defined as

\begin{equation}
C_{ij} \equiv \langle \vec{x}_{i} \cdot \vec{x}_{j} \rangle =
\sum_{\mu} \tilde{M}^{-1}_{ij,\mu\mu}
\label{eqn:cij}
\end{equation}

\noindent In ordinary MD simulations, the thermodynamic average in
(\ref{eqn:cij}) is replaced with the time average taken over the
simulated trajectory (ergodicity assumption). Due to the fact that the
$C$ matrix is obtained from $\tilde{M}^{-1}$ after a summation over
the cartesian components, the linear size of $C$ is equal to the
number of protein residues, $N$, instead of $3N$ as for
$\tilde{M}^{-1}$. This ten-fold reduction of information greatly
simplifies the identification of significant correlations between
residues motion.

We conclude this section by discussing a technically important
point. The inversion of the $\tilde{M}$ matrix used in
eqns. (\ref{eqn:fullcij}--\ref{eqn:cij}), as well as a correct
interpretation of equation (\ref{eqn:boltz}) are possible only within
the subspace orthogonal to the eigenvectors of $\tilde{M}$ associated
with zero eigenvalues. Physically this corresponds to omit the
structural modifications that cost no energy (zero modes). Due to the
invariance of Hamiltonian (\ref{eqn:ham}) under rotations and
translations of the Cartesian reference frame, there will always be at
least six zero modes. This number can, however, be larger if the
$\tilde{M}$ matrix is sparse.The presence of additional spurious modes
in Gaussian network models that incorporate only $C_\alpha$
coordinates is usually achieved by two means: either a reduction of
the dimensionality of $\tilde{M}$ (as in GNM) or by using large
interaction cutoffs in the range 10-15 \AA (as in ANM). The model
discussed here allows to use physically-appealing interaction cutoffs
of the order of 7 \AA, as for GNM, and yet retaining the full
three-dimensional detail in the $\tilde{M}$ matrix. As will be shown
later, these ingredients are necessary to capture the finer aspects of
protein vibrations such as the correlation of residues' motion, while,
consistently with previous studies, the overall mobility of individual
residues is rather insensitive to the details of the model
\cite{coarseanm02a,coarseanm02b}.\\ In fact, we found that GNM, ANM
and $\beta$GM have a similar performance on the prediction of
experimental B-factors. This was established using high-resolution,
single-chain proteins taken from the non-redundant pdb-select list
\cite{pdbselect}. We restricted to proteins length between 50 and 200
residues and excluded from the comparison the first and last 5
residues to avoid biases due to enhanced terminal mobility. Overall we
selected 36 proteins determined with Xray and 31 with NMR.  For
simplicity we summarise the level of agreement as the average of the
non-parametric rank correlation, $\tau$ \cite{halle2002}. This
analysis does not rely on the knowledge of the probability
distribution from which the points (pairs of data) are taken.  What
matters is the agreement of the ranking of the points according to
each of the two variables.  In case of perfect [anti]-correlation the
Kendall parameter $\tau$ takes on the value 1 [-1] and usually provides a more
stringent (and robust) measure than linear correlation\cite{NR}.  For
the X-ray set and using GNM, $\tau$ ranged from 0.37 to 0.39 for
cutoffs in the range 7.5 - 15 \AA. For the same range $\beta$GM gave $
0.34 < \tau < 0.37$, while for ANM $ 0.30 < \tau < 0.37$ for
interaction ranges 10-15 \AA. The comparison against NMR
temperature-factors provided higher correlations as already observed
in ref. \cite{normod}. For the same cutoff ranges reported above one
has for GNM: $ 0.45 < \tau < 0.47$, for $\beta$GM: $ 0.46 < \tau <
0.48$ and for ANM $ 0.42 < \tau < 0.48$.

In summary, the novel model discussed here allows to incorporate, in
an effective Hamiltonian, not only backbone-backbone interactions but
also backbone-sidechain and sidechain-sidechain ones. The sidechain
degrees of freedom are entirely controlled by the $C_\alpha$
positions. This has the crucial implication that the computational
effort required to characterize the vibrational properties of the
system is exactly the same as for models that incorporate only
interactions between $C_\alpha$ pairs.

\section{Results and Discussion}

In this section we shall examine the extent to which suitable
topology-based harmonic models can capture the details of the
near-native vibrations of proteins in thermal equilibrium. Given the
present impossibility to probe experimentally the various
thermodynamical quantities discussed before, it is mandatory to choose
as a reference the results of an all-atom molecular dynamics
calculation performed on the HIV-1 protease, in complex with TIMMNR
peptide model substrate \cite{piana2002,piana2002b}. All MD dynamical
averages were calculated after discarding an initial interval of a few
ns over which the protease complexed to the model substrate was
equilibrated \cite{piana2002,piana2002b}. The configuration obtained
of the end of the equilibration protocol was taken as the reference
structure for the Gaussian approach. The inversion of the symmetric
$\tilde{M}$ matrix, necessary to characterize the system equilibrium
dynamics, was done exploiting the Householder reduction \cite{NR} and
took about 10 minutes on a personal computer.

For the comparison, it is important to remark that although molecular
dynamics studies can reproduce reliably a variety of experimental
quantities they ultimately rely on empirical potentials which may be
imperfectly parametrised. Besides this issue, it should also be noted
that, usually, it is not easy to ascertain whether the simulated
trajectory is sufficiently long that thermodynamic averages as in
eq. (\ref{eqn:cij}) can be legitimately replaced with dynamical ones,
though a recent study has indicated some valuable criteria for this
purpose \cite{Hess2002}. This potential limitations of general MD
approaches should therefore be borne in mind also in the present
context.
 
The series of tests carried out to ascertain the consistency among MD
results and the one of gaussian models include the comparison of
temperature factors, covariance matrices and essential subspaces. The
findings, summarised below, provide a direct and strong indication
that the $\beta$GM is apt for capturing several aspects of proteins'
near-native vibrational dynamics with an accuracy that rivals with
techniques based on all-atom potentials.

The generalised energy function of eq. (\ref{eqn:ham}) contains
several parameters; one of these, the interaction amplitude of
$C_\alpha$ pairs, $V^{\prime \prime}_{CA-CA}$, can be conveniently
taken as the energy unit. The other parameters are the interaction
cutoff, $R$ (which enters in the definition of the contact matrices)
and the amplitudes of the $C_\beta$-$C_\beta$ and $C_\beta$-$C_\alpha$
interactions as well as the extra strength of the peptide term, $k$.
For reasons of simplicity the strength of these interactions have been
chosen of the same order as $V^{\prime \prime}_{CA-CA}$= $k=1$,
$V^{\prime \prime}_{CA-CB}=V^{\prime \prime}_{CB-CB}=1/2$. This
choice was done for reasons of simplicity but is not particularly
restrictive due to the fact that effective interactions between
sidechains are expected to be of the same order as
$C_\alpha$-$C_\alpha$ ones \cite{stabloc}; in addition the precise
value of $k$ will mostly impact on the high frequency vibrational
spectrum of the system. Hence, it can be anticipated that the system
elastic response should mostly depend on the value of the interaction
radius, $R$, which has been accordingly varied in our analysis.

We first discuss the possibility to predict the temperature factors
encountered in molecular dynamics. This type of validation has been
considered before by Doruker {\em et al.} in connection with the
anisotropic gaussian model and reported a good consistency with
dynamical simulations \cite{doruker2000}.

\begin{figure}
\includegraphics[width=3.0in]{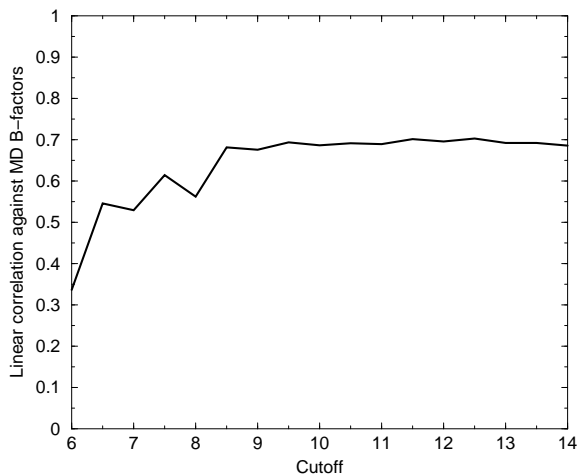}
\caption{Linear correlation coefficient for the temperature factors of
the 204 residues in the HIV-1 PR/SUB complex.}
\label{fig:bfact}
\end{figure}

As a measure of the agreement between the residues mean square
fluctuations observed in MD and those predicted from the Gaussian
models, see eqn. (\ref{eqn:bfact}), we considered the linear
correlation coefficient. The degree of correlation as a function of
the interaction cutoff radius for $\beta$GM is shown in Fig. \ref{fig:bfact}

\noindent The performance of the $\beta$GM particularly stable beyond
an interaction cutoff of about 8.0 \AA. Given the large number of
residues in the system (198 for the protein and 6 for the substrate)
over which the B-factors are calculated, it is certainly possible to
conclude that the correlation coefficient approaching 0.7 visible in
Fig. \ref{fig:bfact} is statistically significant. A more quantitative
assessment of the statistical significance could be done using the
Student's $t$-test or related methods \cite{NR}, although such
analysis usually rely on the assumption that the joint distribution of
the correlated variables is binormal (which is not necessarily
satisfied for B-factors).  An alternative way is to recourse to the
non-parametric Kendall test mentioned before \cite{NR}, as recently
proposed by Halle \cite{halle2002}.  The Kendall correlation
coefficient among the B-factors of the simulation and those of the $\beta$
Gaussian model (for $R \approx$ 7.5 \AA) is $\tau \approx 0.61$ and
amply satisfies all ordinary criteria for statistical significance.
 
The successful comparison of the B-factors confirms the general
agreement between the overall residues' motion in MD and the
equilibrium dynamics predicted by Gaussian models\cite{doruker2000};
in fact, for the same cutoff used above ang agsinst the same MD data
of the HIV-1 PR/SUB complex, GNM provides a linear correlation
coefficient of 0.61 while the Kendall parameter $\tau$ is equal to
0.59. However, the finer details of such accord have not, to the best
of our knowledge, been explored yet and hence become the focus of our
subsequent analysis which is based on a comparison of the MD and
$\beta$GM covariance matrices. This test is particularly important due
to the wealth of biological and chemical information that can been
extracted from the covariance (essential dynamics) analysis
\cite{Amadei93,garcia92,brooks2003}.

In the context of HIV-1 Pr
\cite{condra,patick,BIOCH88,gulnik,hiv1,apr,condra1,boucher1,Molla,Marko}
these motions have a direct mechanical bearing on the structural
modulation of the active site, even though they are located remotely
from it\cite{piana2002,piana2002b,pcr}.

As a first case we consider the reduced covariance matrices. To allow
a straightforward comparison of the theoretical and numerical results
rather than working directly in terms of the reduced covariance
matrix, it is useful to focus on the normalised version, which is
dimensionless:

\begin{equation}
\tilde{C}_{ij} = {\langle \vec{x}_{i} \cdot \vec{x}_{j} \rangle \over
\sqrt{\langle |\vec{x}_{i}|^2 \rangle\, \langle |\vec{x}_{j}|^2
\rangle}}
\label{eqn:cijnorm}
\end{equation}

\begin{figure}
\includegraphics[width=3.0in]{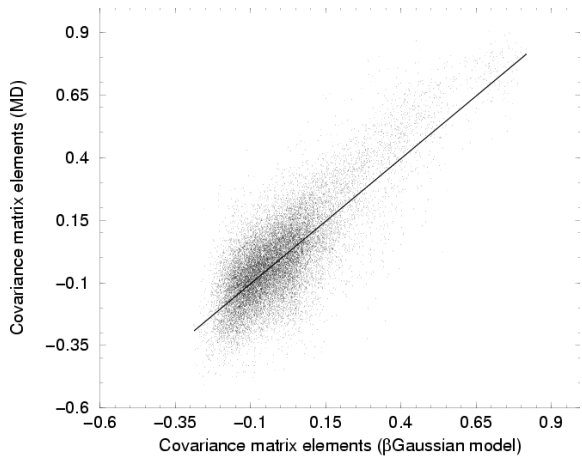}
\caption{Scatter plot of corresponding entries of the covariance
matrices obtained within the $\beta$ Gaussian model ($R=7.5$ \AA) and
from the 14ns MD simulation on the HIV-1 PR/SUB complex. The linear
correlation coefficient over the  $2\cdot 10^4$ data points is 0.80 .}
\label{fig:cov2}
\end{figure}

\noindent The scatter plot of Fig. \ref{fig:cov2} summarises the
degree of accord between the normalised covariance matrix of the MD
simulation and that of the $\beta$-Gaussian model, for a cutoff 0f 7.5
\AA. The number of entries in the plot is about $2\cdot 10^4$, equal
to the number of distinct entries in the 204x204 $\tilde{C}_{ij}$
matrix.  To avoid introducing artificial biases in the correlation,
the diagonal elements of the normalised matrices (which are all equal
to 1) have been omitted from the plot.

The linear correlation coefficient among the two sets of data is 0.80
(and is stable in the neighborhood of $R=7.5$ \AA).  We do not attempt
to provide a quantitative measure for the statistical significance of
the linear correlation in Fig. \ref{fig:cov2}. In fact, on one hand,
as visible in Fig. \ref{fig:cov2}, the joint distribution for
covariance matrix elements is only approximately binormal, and hence
the traditional tests of linear regression significance are of dubious
applicability. On the other, the Kendall correlation measure requires
to consider all possible pairs of points in the scatter plot: this
makes the analysis impractical and disproportionate to the main goal
which is to ascertain the existence of the accord between topological
gaussian models and MD results. In fact, rather than measuring the
correlation in absolute terms, we shall compare the accord of gaussian
models and MD simulations against the degree of ``internal''
consistency of the simulated dynamical trajectory itself.

We conclude the discussion of the scatter plot of Fig. \ref{fig:cov2}
by mentioning that, in case of perfect correlation of two
$\tilde{C}_{ij}$ sets, due to the normalization condition of
eqn. \ref{eqn:cijnorm}, the data would align along the diagonal of the
graph in Fig. \ref{fig:cov2}. Interestingly, despite the scatter
visible in the same figure, the interpolating line lies very close to
the diagonal, having a slope of $s=0.97$ (taking the $\beta$GM
covariance as the independent variable). This fact is useful in
illustrating the effects of the cutoff, $R$, on the accord with MD
covariance elements. While for $R=10$\AA the slope is still good,
$s=1.05$, it deteriorates for $R = 15$\AA, where $s=1.64$. This effect
is even more pronounced if the $C_\beta$'s are not included in the
model. For example, for $R = 15$ \AA\ the observed slope was $s =
3.41$.

The quantification of the self-consistency of MD dynamical
trajectories is an extremely important issue since it can provide an
{\em a posteriori} indication of whether the dynamical sampling of the
phase-space was sufficiently to obtain reliable thermodynamic
averages. The analysis usually starts from the calculation of two
covariance matrices pertaining to the first and second halves of the
MD trajectory.

Ths two matrices could then be compared entry by entry, as done above.
However, the most appropriate procedure is not to compare the
corresponding matrix elements, but rather the physically-important
(essential) eigenspaces, that is the linear spaces spanned by the
eigenvectors of $M^{-1}$ associated to the largest eigenvalues. A
number of studies have suggested ways of measuring this consistency.

The first method of comparison that we will be taken into account is
the one introduced by Amadei {\em et al} \cite{Amadei99} which focuses
on the top $n$ eigenvectors of the covariance matrices under
comparison. These eigenvectors describe the most significant modes of
vibration of the molecule in the three-dimensional space. We stress
here that the covariance matrix considered here is not the reduced one
of eq. (\ref{eqn:cij}), whose size is $N$x$N$, but is the full one (the
$M^{-1}$ matrix) of size $3N$x$3N$ that contains the three-dimensional
information about correlated motion of pairs of residues, see
e.g. eqn. (\ref{eqn:fullcij}).

By denoting the top $n$ eigenvectors of the two matrices under
comparison as $\{\vec\eta\}$ and $\{\vec\nu\}$, the degree of overlap
of the essential subspaces is defined as the root mean square inner
product (RMSIP) of all pairs of eigenvectors in the two sets
\cite{Amadei99}:

\begin{equation}
RMSIP = \sqrt{ {1 \over n} \sum_{i,j} | \vec\eta_i \cdot \vec\nu_j
|^2}
\label{eqn:amadei}
\end{equation}

\noindent Customarily, the analysis is restricted to the top $n=10$
eigenvectors. We have measured the RMSIP in eqn.  (\ref{eqn:amadei})
when $\{\eta\}$ and $\{\nu\}$ come from the essential subspaces of the
first and second halves of the 14ns MD trajectory of the HIV-1 PR/SUB
complex. The calculated RMSIP value, eq. \ref{eqn:amadei}, was
0.71. Amadei et al. \cite{Amadei99} have also proposed a series of
approximate tests to ascertain the statistical relevance of the
observed overlap. Based on their analysis, we can conclude that the
value obtained here, for the given system size of 204 amino acids, has
a probability to have arisen by chance that is, by far, inferior to
the conventional threshold of 1 \%. This supports the fact that the MD
trajectory was sufficiently long to contain significant physical
information about the system equilibrium dynamics.

Having in mind the level of internal consistency of the present
reference MD trajectory we have turned to measuring the RMSIP between
the essential spaces of the whole dynamical trajectory and those of
the $\beta$ Gaussian model.

\begin{figure}
\includegraphics[width=3.0in]{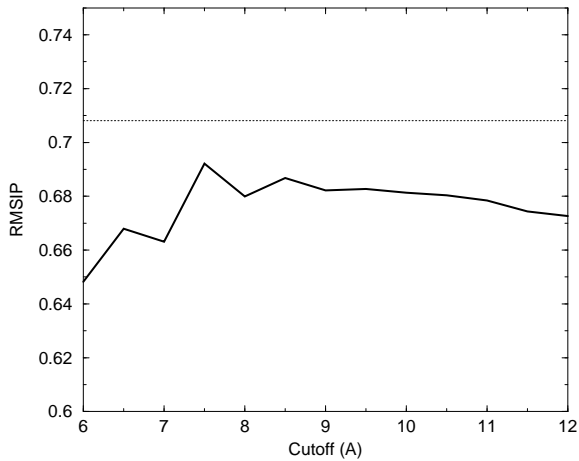}
\caption{Degree of correlation (root mean square inner product), see
eqn. \protect(\ref{eqn:amadei}), of the essential subspaces of the MD
simulation and of the gaussian models as a function of the interaction
cutoff, $R$. The thick curve denotes the performance of the $\beta$
Gaussian model, while the horizontal dotted line indicates the overlap
of the essential subspaces of the first and second halves of the MD
trajectory.}
\label{fig:amadei}
\end{figure}

The resulting trend for the subspaces overlap is shown in
Fig. \ref{fig:amadei} as a function of the interaction cutoff,
$R$. The best performance of the model is obtained for a cutoff of $R
\approx 7.5$ \AA.  The corresponding overlap value of 0.68 is very
close to the internal overlap of the MD trajectory. We wish to remark
that such values of RMSIP are highly non-trivial due to the large size
of the full covariance matrix (612x612). This implies that the
probability to observe a given overlap, $q$, between two random unit
vectors decreases extremely rapidly as $q$ approaches 1
\cite{Amadei99}. As a consequence, the MD simulation time required to
reach a given target value for the internal RMSIP consistency,
$\bar{q}$, grows very repidly with $\bar{q}$.

This observation clarifies the utility of the Gaussian approach. With
a modest computational investment, required by the diagonalization of
a $3N$x$3N$ matrix, one obtains a description of the protein essential
dynamics that,within a molecular dynamics framework where all atom
effective potentials are used, requires a considerably heavier
computational investment. Further improvements over the $\beta$GM
performance are obviously possible within MD, but at the price of a
rapidly growing computing time.

The usefulness of the $\beta$GM is further supported by yet another
type of analysis, which concludes the series of tests carried out
here. This last approach follows a more precise measure for the
agreement of the essential subspaces recently introduced by Hess
\cite{Hess2002}. The new measure is an improvement over the definition
of eqn. (\ref{eqn:amadei}) since it removes both the subjectivity of
the choice of $n$, assigns more importance to the physically-relevant
eigenspaces and deals correctly with the presence of spectral
degeneracies.

\begin{figure}
\includegraphics[width=3.0in]{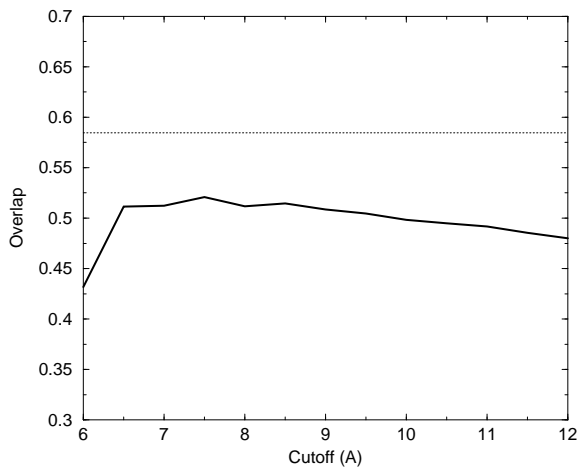}
\caption{Degree of overlap according to the measure introduced by Hess
\protect{\cite{Hess2002}} of the essential subspaces of the MD
simulation and of the gaussian models as a function of the interaction
cutoff, $R$. The thick curve denotes the performance of the $\beta$
Gaussian model, while the horizontal dotted line indicates the overlap
of the essential subspaces of the first and second halves of the MD
trajectory.}
\label{fig:berkcor}
\end{figure}

Unlike the measure of Amadei {\em et al.}, the one of Hess is
sensitive to the actual values of the eigenvalues of $M^{-1}$. Since
the energy units of the gaussian models considered here has been
chosen arbitrarily, a proper normalisation of the spectrum of the
covariance matrices has to be carried out in order to use the measure
of Hess for the comparison against MD. For this reason, the degree of
overlap of the matrices was carried out after having uniformly
rescaled the eigenvalues of $M^{-1}$ so that the trace of $M^{-1}$ was
equal to 1 for both systems. Physically this corresponds to a
normalization of the average residues mean square displacement.

The results for this analysis are shown in
Fig. \ref{fig:berkcor}.  The enhanced stringency of this test, which
is extended to the whole vibrational spectrum, and not just to the top
10 modes, is reflected in an overall decrease of the overlap with
respect to Fig. \ref{fig:amadei}. This finer measure also reflects
better than RMSIP the higher degree os inner consistency of the MD
trajectory, as opposed to the MD-$\beta$GM accord. Although the best
reference for our model would be provided by a simulation run
sufficiently long that the inner MD overlap approaches 1, this is not
presently feasible due to the slow (approximately logarithmic)
increase of the inner overlap with the length of the MD run. Despite
this fact, the results fully confirm the previous conclusions namely
that the $\beta$ Gaussian model can predict, with good statistical
confidence, equilibrium dynamical properties of proteins and, in
particular, identify the relevant modes of vibration of the system.

\begin{figure}
\includegraphics[width=3.0in]{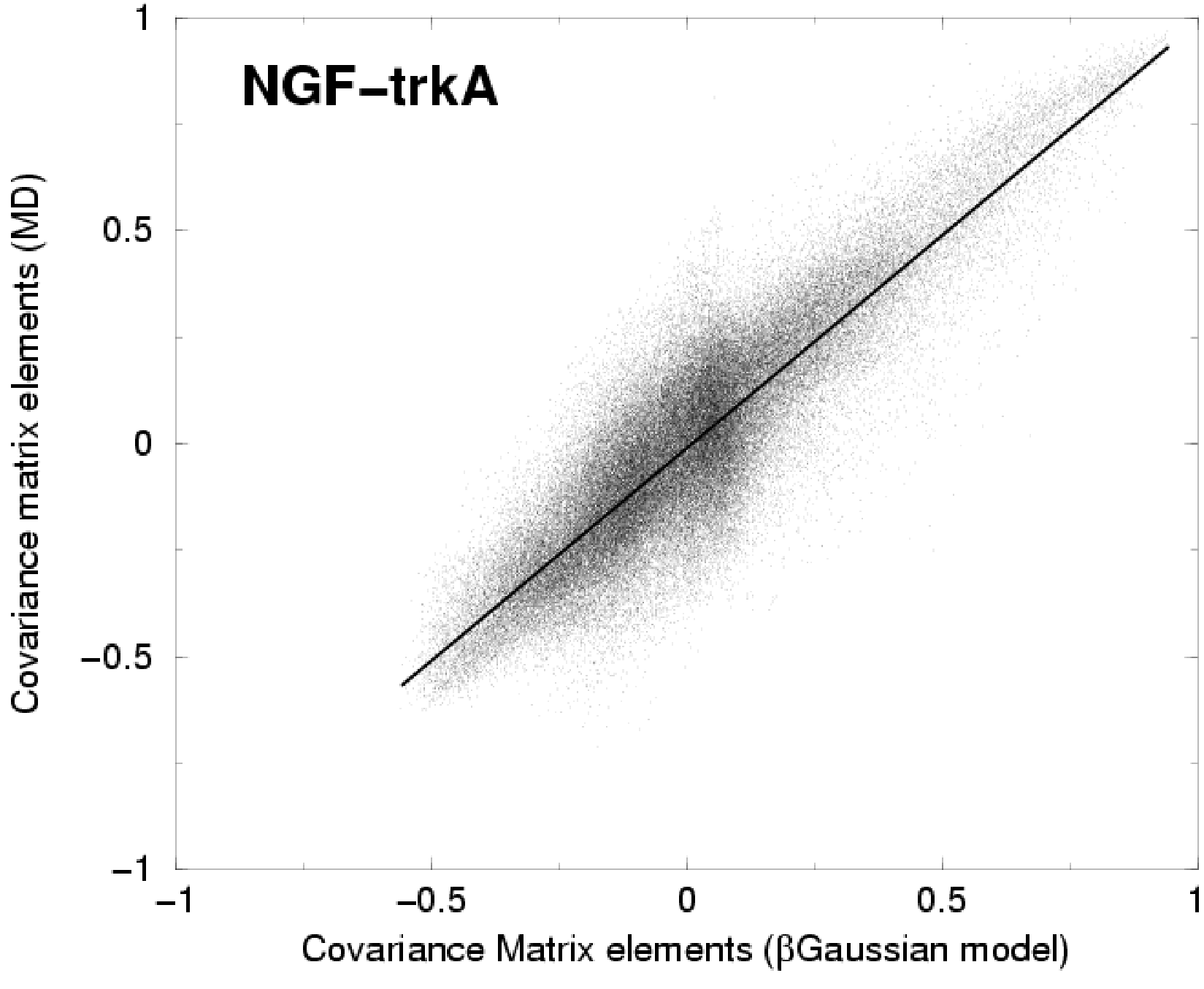}
\caption{Scatter plot of corresponding entries of the covariance
matrices obtained within the $\beta$ Gaussian model ($R=7.5$ \AA) and
from the 2.6ns MD simulation on the NGF-trkA complex
\protect\cite{settanni_ngf}. The linear correlation coefficient over
the nearly $10^5$ data points is 0.86}
\label{fig:ngf}
\end{figure}

\noindent The reference system used here for the comparison between the
model and MD simulations was chosen for both its biological
significance and for the availability of MD data collected over the
rather long simulation time.  The $\beta$ Gaussian model is, however,
of general applicability, and to confirm the robustness of the
strategy we have considered another biologically important reference
system, the NGF-trkA complex.  This is constituted by a protein dimer,
the nerve growth factor (NGF), complexed with the tyrosine kinase A
receptor (trkA); altogether the system comprises 431 residues. The
dynamics of the complex in aqueous solution was recently simulated for
a time span of 2.6 ns using all-atom potentials \cite{settanni_ngf}.
The resulting normalised covariance matrix was compared with the one
obtained from the $\beta$ Gaussian model. The linear correlation
coefficient over the nearly $10^5$ corresponding distinct entries of
the matrices is 0.86, as visible in Fig. \ref{fig:ngf}.

\noindent This result confirms the viability of the Gaussian approach
to capture the details of the large-scale protein movements.  However,
due to the simplicity of the model interaction potentials (all pairs
of $C_\alpha$'s and $C_\beta$'s interact with the same strength) one
may foresee that the model is not suitable for modelling the
vibrational dynamics of proteins where electrostatic effects or
disulfide bonds play an important role for native stability or
funtionality.

\vskip 0.5cm

{\em Biological implications}

\begin{figure}
\includegraphics[width=3.0in]{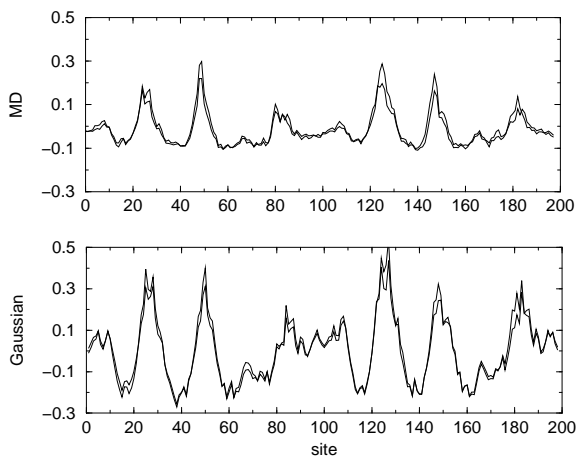}
\caption{The two curves in each panel indicate the degree of
correlation of the motion between the two central atoms of the
substrate and the 198 residues in the subunits of the HIV-1
protease. The top panel reports the MD findings, while the bottom one
pertains to the $\beta$ Gaussian model.}
\label{fig:dimercontour}
\end{figure}

In the MD calculations by Piana et al.\cite{piana2002,piana2002b,pcr}
the essential dynamics analysis has allowed to identify the sites
that, despite being spatially distant from the active site have a
strong mechanical influence on the structural modulation of the
regions binding the substrate. We now compare the findings obtained
with MD with those obtained with our model.  The degree of mechanical
coupling observed in the MD trajectory between the substrate motion
and the HIV-1 protease subunits is visible in the top panel of Fig.
\ref{fig:dimercontour}. The two curves in the plot represent the
profile of the reduced covariance matrix, $C_{ij}$, between the two
central atoms of the peptide and the 198 protease residues.
Interestingly, the regions that correlate significantly with the
substrate motion are those that have been indicated as rather
under-constrained by studies where the theory of rigidity has been
applied to characterize the enzyme elasticity \cite{thorpe2001}.  The
two facts provide a consistent picture for the HIV-1 PR mechanics
since any functionally-relevant mechanical coupling must intuitively
involve mobile (and hence under-constrained)
regions. The identification of the mobile regions of
HIV-1 protease has also been previously addressed with Gaussian
network models in a series of studies which also allowed to identify
the residues important for protein stability \cite{bah98,hivgnm03} or
otherwise taking part to crucially important networks of key native
contacts\cite{hiv-gauss}.

The bottom panel of figure \ref{fig:dimercontour} represents the
corresponding correlation profiles calculated within the $\beta$GM for
a cutoff $R= 7.5$ \AA (which we take as an optimal value from previous
analysis). The degree of agreement of the profiles across the two
panels is remarkable and, the main difference appears to be due to an
overemphasis of negative correlations in the gaussian profiles.

In both sets of data one observes a strong positive correlation
between the substrate motion and the regions 24-30 and 45-55. This
direct mechanical coupling is of immediately interpreted due to the
fact that the first region comprises the cleavage site while the
second involves the tips of the protease flaps.

The essential-space analysis of the MD trajectory revealed that the
two regions embrace the substrate and involve it in a rotational
``nutcracker-like'' motion. As a consequence of this rotation, the
regions near the flaps elbows, 37-41 and 61-73, undergo a
counter-movement that results in a negative correlation with the
substrate motion. This effect is clearly visible in the gaussian
profiles of Fig. \ref{fig:dimercontour}, while it is less pronounced,
but still significant, in the MD results.

As remarked in the Theory section, the intimate connection between the
linear response theory and the covariance matrix allows to conclude
that a force applied in correspondence of sites around residues 40 and
63 should affect the protease-substrate coupling. This effect was
indeed observed in the MD simulation where the motion of the substrate
towards the cleavage site was strongly affected by the constraints on
these regions (see Fig. 6 in ref. \cite{piana2002}).

 It is by virtue of this mechanically-important coupling that it is
possible to rationalize the emergence of mutations causing drug
resistance in correspondence of sites far from the cleavage region
(e.g. M63I, M46I-L and L47V).  In fact, as the explicit MD calculation
has shown, the high degree of such coupling between such sites and the
cleavage is such that the detailed chemical identity of the former
strongly influence the substrate binding affinity of the latter. In
particular, the mutations observed in clinics
\cite{apr,condra1,boucher1,Molla,Marko} are arguably the result of a
chemical fine-tuning that retains the native functionality of the
enzyme while decreasing its affinity for inhibiting drugs.

The details of how the enzymatic reaction kinetics changes upon amino
acid mutations is beyond the reach of the topological models presented
here. The gaussian scheme, in fact, is entirely adequate to identify
which sites influence mechanically the active site motion but, due to
the fact that all amino acids (except for GLY which lacks the
$C_\beta$ centroid) are treated equally we cannot explore the
ramifications of changing the amino acid identity into the
cleavage-region and substrate coupling. The effect of one of such
mutations (M46I) on substrate motion has instead been fully taken into
account in ref. \cite{piana2002b}.
Further improvements of the $\beta$ Gaussian model may be possible by
optimizing the distance of the $C_\beta$ centroids from their
respective $C_\alpha$'s (so to better capture the displacement of the
sidechains centres of mass). The model could also be extended to
include, at the simplest possible level, the effects of thermal
denaturation through a self-consistent temperature-dependent weakening
of the strength of the harmonic couplings, as done in
refs. \cite{gaussian,normod}.

\section{Conclusions}

We have examined the extent to which the dynamical properties of a
protein in thermodynamic equilibrium can be accounted for through
solvable models. The starting proint of our analysis is the quadratic
approximation of the free energy landscape in terms of the deviations
of amino acids from their reference positions in the known native
state.  We have adopted a novel description of the amino acids which
allows to consider the presence of effective $C_\alpha$ centroids
whose degrees of freedom are entirely controlled by the $C_\alpha$
atoms. As a result, the model that is considered, is able to account
for the directionality of amino acid sidechains while retaining the
same degree of complexity as models based on $C_\alpha$ representation
only. Various equilibrium quantities apt to characterize the most
relevant modes of vibrations of proteins are considered. In
particular, we focussed on (in increasing order of complexity and
detail) the B-factors, the covariance matrices and the
essential dynamical subspace. Our results have been compared against
the analogous quantities obtained through a 14ns molecular dynamics
simulation carried out on the HIV-1 PR enzyme in complex with a TIMMNR
peptide 
substrate.

As fas as overall equilibrium dynamical properties are concerned, the
$\beta$ Gaussian model provides a picture that is in remarkable
agreement with the MD results. In fact, the essential subspace
predicted theoretically appears to have a degree of consistency with
MD results that is close to the ``inner consistence'' of a 14ns MD
simulation with itself.

This provides a strong indication that suitable quadratic models can
provide a powerful and accurate tool for characterizing the
vibrational motions of proteins near their native state while
requiring only a modest investment of computational resources.  Other
important properties of protein dynamics and functionality that
strongly depend on the sequence composition or on out-of-equilibrium
conditions are, at the moment, beyond the reach of such simplified
approaches considered here. For this reason we believe that the
Gaussian approach would be ideally used in conjunction with molecular
dynamics by providing, prior to investing significant computational
resources in all-atom simulations, a fast but accurate
characterization of a protein's near-native motion.

\section{Acknowledgments}

We are indebted to Stefano Piana, Michele Cascella, Giorgio Colombo,
Paolo De Los Rios, Gianluca Lattanzi, Luca Marsella and Gianni
Settanni for useful suggestions and advice. We acknowledge financial
support from INFM and Cofin MIUR 2001.

\end{document}